\documentclass[12pt]{article}
\usepackage{amsfonts,amssymb,epsfig,cite}

\begin{document}

%% \date{February 26, 2002}

\def\reff#1{(\ref{#1})}
\newcommand{\be}{\begin{equation}}
\newcommand{\ee}{\end{equation}}
\newcommand{\<}{\langle}
\renewcommand{\>}{\rangle}
\newcommand{\CITE}{\cite}

%%%  \ltapprox and \gtapprox produce > and < signs with twiddle underneath
\def\spose#1{\hbox to 0pt{#1\hss}}
\def\lesssim{\mathrel{\spose{\lower 3pt\hbox{$\mathchar"218$}}
 \raise 2.0pt\hbox{$\mathchar"13C$}}}
\def\gtapprox{\mathrel{\spose{\lower 3pt\hbox{$\mathchar"218$}}
 \raise 2.0pt\hbox{$\mathchar"13E$}}}

\def\bsigma{\mbox{\protect\boldmath $\sigma$}}
\def\bpi{\mbox{\protect\boldmath $\pi$}}
\def\smfrac#1#2{{\textstyle\frac{#1}{#2}}}
\def\smhalf{ {\smfrac{1}{2}} }

\newcommand{\re}{\mathop{\rm Re}\nolimits}
\newcommand{\im}{\mathop{\rm Im}\nolimits}
\newcommand{\tr}{\mathop{\rm tr}\nolimits}
\newcommand{\fr}{\frac}
\newcommand{\diti}{\frac{\mathrm{d}^2t}{(2 \pi)^2}}
\newcommand{\bz}{{\mathbf 0}}

\def\Z{{\mathbb Z}}
\def\R{{\mathbb R}}
\def\C{{\mathbb C}}

\title{Third Virial Coefficient for 4-Arm and 6-Arm Star Polymers}

\author{ Sergio Caracciolo \\
Dipartimento di Fisica, Universit\`a degli Studi di Milano,  \\
  and INFN -- Sezione di Milano I \\
  Via Celoria 16, I-20133 Milano, Italy \\
  e-mail: {\tt Sergio.Caracciolo@mi.infn.it} \\[0.2cm]
Bortolo Matteo Mognetti \\
Institut f\"ur Physik, Johannes Gutenberg-Universit\"at, \\
  Staudinger Weg 7, D-55099 Mainz, Germany \\
  e-mail: {\tt mognetti@uni-mainz.de} \\[0.2cm]
Andrea Pelissetto \\
  Dipartimento di Fisica and INFN -- Sezione di Roma I  \\
  Universit\`a degli Studi di Roma ``La Sapienza'' \\
  P.le A. Moro 2, I-00185 Roma, Italy \\
  e-mail: {\tt Andrea.Pelissetto@roma1.infn.it }}

\maketitle
\thispagestyle{empty}   % Suppress page number on front page.

\begin{abstract}

We discuss the computation of the third virial coefficient in polymer
systems, focusing on an additional contribution absent in 
the case of monoatomic fluids. We determine the interpenetration 
ratio and several quantities that involve the third virial
coefficient for star polymers with 4 and 6 arms in the good-solvent
regime, in the limit of a large degree of polymerization.

% 61.25.Hq  Macromolecular and polymer solutions; polymer melts; swelling 
% 82.35.Lr  Physical properties of polymers 

\end{abstract}

\clearpage

\section{Introduction}

In the dilute regime the osmotic pressure of a polymer solution 
can be predicted successfully by using the 
virial expansion, which we write as  
\begin{equation}
Z \equiv   {M \Pi \over R T \rho} = {\Pi\over k_B T c} = 
1 + \sum_{n=1} B_{n+1} c^n,
\label{expanZ1}
\end{equation}
where $c$ is the polymer number density, $\rho$ the weight concentration,
$M$ the molar mass of the polymer, $T$ the absolute
temperature, and $k_B$ and $R$ the Boltzmann and the ideal-gas constant,
respectively.
The virial coefficients $B_n$ depend on the degree of polymerization $N$ and 
on the chemical details. However, in the good-solvent regime
renormalization-group arguments\cite{deGennes-79,Freed-87,Schaefer-99}
indicate that, for $N\to \infty$, the ratios
\be 
A_{n+1} \equiv  B_{n+1} \hat{R}_g^{-3n},
\label{expanZ}
\end{equation}
where $\hat{R}_g$ is the zero-density radius of gyration, 
approach universal constants $A_{n+1}^*$ that are independent of chemical
details and depend only on the polymer large-scale structure.

Much numerical and experimental work has been devoted to the calculation of the 
second virial coefficient $B_2$. Results for the higher-order coefficients 
are instead rare, both experimentally and numerically. In recent years some
numerical computations of the third osmotic virial coefficient for 
solutions of polymers of different architecture have been reported.
\cite{Bruns-97,Vega-00,VLMDS-00,SOKN-02,PH-05,CMP-vir,SKOKN-07,CMP-TPM}
However, in essentially all works an incorrect expression for the 
third virial coefficient was used. The correct expression, which is 
valid for any fluid of flexible molecules, was derived in ref.~\CITE{CMP-vir},
and used to determine the universal ratio $A_3^*$ for linear polymers in
the good-solvent regime. Let us report here the result.
Let us consider a molecular fluid in which each molecule is formed by $N$ units 
interacting by means of an intramolecular potential 
$V_{\rm intra}({\mathbf r}_1,\ldots, {\mathbf r}_N)$, 
where ${\mathbf r}_1,\ldots, {\mathbf r}_N$ are the unit positions. 
Molecules $i$ and $j$ interact by means of an intermolecular potential
$V_{\rm inter}$ that depends on the positions $\{{\mathbf r}^{(i)}_a\}$
and $\{{\mathbf r}^{(j)}_a\}$. 
Given a quantity ${\cal O}$ which depends on the coordinates of two molecules,
we define a zero-density average $\<\cdot\>^0$ as 
\begin{eqnarray}
&& \langle {\cal O}\rangle^0_{{\mathbf r}^{(1)}, {\mathbf r}^{(2)}} \equiv
  {1\over Q_2}
  \int d{\mathbf r}_2^{(1)}\ldots d{\mathbf r}_N^{(1)}
       d{\mathbf r}_2^{(2)}\ldots d{\mathbf r}_N^{(2)}\, 
        {\cal O} \, 
 \exp [-\beta V_{\rm intra}(\{{\mathbf r}^{(1)}\})
       -\beta V_{\rm intra}(\{{\mathbf r}^{(2)}\}) ], \nonumber \\
&& Q_2 \equiv \int d{\mathbf r}_2^{(1)}\ldots d{\mathbf r}_N^{(1)}
       d{\mathbf r}_2^{(2)}\ldots d{\mathbf r}_N^{(2)}\,
 \exp [-\beta V_{\rm intra}(\{{\mathbf r}^{(1)}\})
       -\beta V_{\rm intra}(\{{\mathbf r}^{(2)}\}) ].
\end{eqnarray}
The meaning of this average is easily understood: we fix the position of the 
first unit of the two molecules to avoid irrelevant volume factors 
and average over all possible conformations, weighting each conformation
with the intramolecular Hamiltonian only (which correspond to consider the 
zero-density limit). Analogously, 
given a quantity ${\cal O}$ that depends on the coordinates of three molecules,
we define an average $\langle {\cal O}
\rangle^0_{{\mathbf r}_1^{(1)}, {\mathbf r}_1^{(2)},{\mathbf r}_1^{(3)}}$: 
it corresponds to averaging over all possible conformations of the three 
molecules keeping the first unit of the three molecules fixed in 
${\mathbf r}_1^{(1)}, {\mathbf r}_1^{(2)},{\mathbf r}_1^{(3)}$.
In terms of these quantities we define
\begin{eqnarray}
I_2 &\equiv & \int d^3 {\mathbf r}_{12} \, 
       \langle f_{12} \rangle^0_{{\mathbf 0},{\mathbf r}_{12}} ,
\\
I_3 &\equiv & 
    \int d^3{\mathbf r}_{12} d^3{\mathbf r}_{13}
     \left\langle f_{12}f_{13}f_{23}
     \right\rangle^0_{{\mathbf 0},{\mathbf r}_{12},{\mathbf r}_{13}} ,
\\
T_1 &\equiv & \int d^3{\mathbf r}_{12} d^3{\mathbf r}_{13}
      \left\langle f_{12} f_{13} 
       \right\rangle^0_{{\mathbf 0},{\mathbf r}_{12},{\mathbf r}_{13}}
   -\left[ \int d^3{\mathbf r}_{12}
    \left\langle f_{12} \right\rangle^0_{{\mathbf 0},{\mathbf r}_{12}}\right]^2,
\end{eqnarray}
where $f_{ij} = \exp(- \beta V_{\rm inter}) - 1$ is the Mayer function.
The third virial coefficient is then given by:
\begin{eqnarray}
B_3 &=& - {1\over 3} I_3  - T_1 .
\label{B3def} 
\end{eqnarray}
This expression contains two terms: the first one, proportional to $I_3$,
corresponds to the usual term that gives the third virial coefficient in 
monoatomic fluids. In addition, there is a second term that is not present 
in monoatomic fluids and is related to the flexibility of the 
polymer molecule. This additional term was neglected in 
refs.~\CITE{Bruns-97,Vega-00,VLMDS-00,SOKN-02,PH-05,SKOKN-07}, and thus
the corresponding estimates of the third virial coefficient are incorrect. 
Bruns,\cite{Bruns-97} starting from a general expression given in
Yamakawa's book,\cite{Yamakawa-book}
 derives the correct expression for $B_3$, but then he 
neglects $T_1$ in the numerical calculation, stating incorrectly that it can 
be shown that 
such term vanishes for hard-core systems. 
The derivation of $B_3$ given in ref.~\CITE{CMP-vir} does not give a physical
interpretation to the additional term $T_1$. Here we present
a different derivation that follows the approach of ref.~\CITE{Vega-00}.
It clarifies the physical meaning of $T_1$ 
and explains why this term is necessarily present and non-vanishing.

For linear polymers, 
even though $T_1$ does not vanish, its contribution is small. Indeed,
the results of refs.~\CITE{PH-05,CMP-vir} provide the estimates 
\begin{eqnarray}
&& A_3^* =  9.80 \pm 0.02, \\
&& \widehat{A}_3^* = \lim_{N\to\infty}
  \left(-{1\over 3} I_3 R_g^{-6}\right) = 10.60 \pm 0.04, \\
 && \lim_{N\to\infty} T_1 R_g^{-6} = 0.80 \pm 0.05.
\end{eqnarray}
Thus, the contribution $T_1$ lowers the third virial coefficient only by 8\%. 

In this paper, we extend the calculations of ref.~\CITE{CMP-vir} to 
regular star polymers in which $f$ branches of equal molecular weight
are connected to a single branching unit.\cite{GFH-96,Burchard-99}
This type of polymers is particularly interesting. First, they
have several technological applications.\cite{GFH-96} Second,
they show a quite different behavior depending on the number $f$ 
of branches, interpolating between linear polymers and hard colloids. 
Here we shall focus on the cases $f = 4$ and  $f= 6$ with the purpose of 
investigating the quantitative role of the additional term $T_1$
for a different polymer conformation. For both values of  $f$
we find that the additional 
contribution $T_1$ is small: our results show that, for large $N$, 
$T_1/B_3 \approx 0.065$ and 0.05 for $f=4,6$, respectively. 
Note that the relative importance of $T_1$ decreases as $f$ is increased,
indicating that star polymers become increasingly more rigid as $f$
goes to infinity.

The paper is organized as follows. In Section~\ref{sec2} we give a 
new derivation of the expression (\ref{B3def}). 
In Section~\ref{sec3} we explain the 
model we use and the 
simulation method, while in Section~\ref{sec4} we present our results 
and compare them with the existing literature.

\section{A New Derivation of the Third Virial Coefficient for 
         Flexible Molecules} \label{sec2}

Equation (\ref{B3def}) was obtained in ref.~\CITE{CMP-vir}
by first performing an
activity expansion in the grand-canonical ensemble. A different
derivation is reported in ref.~\CITE{Bruns-97}. None of these two
derivations gives any physical insight on the origin of the term $T_1$
and indeed Bruns \cite{Bruns-97}
concluded incorrectly that $T_1 = 0$. Here we present 
a new derivation that clarifies that $T_1$ vanishes only if 
the molecules are rigid, i.e. if the probability of each conformation is 
density independent.

We consider the general case of 
 molecules that have many different internal conformations,
labelled by an index $\alpha$. Each conformation has a Boltzmann weight 
$p_\alpha$, $p_\alpha \propto \exp(-\beta V_{\rm intra})$, 
normalized so that $\sum_\alpha p_\alpha = 1$. 
For convenience, we assume that the number of conformations is finite,
as it occurs in lattice models, but the results are clearly valid also 
in the general case in which there is an infinite number of conformations
(it is enough to replace sums by integrals).
Particles interact by means of a pairwise potential 
$V(r,\alpha,\beta)$ that depends on the relative distance 
${r}$ (we fix a reference point on each molecule) and on the 
internal conformations. As usual, we introduce the Mayer function
\be
f_{ij}(r,\alpha_i\alpha_j) \equiv
  \exp[- \beta V(r,\alpha_i,\alpha_j)] - 1.
\end{equation}
Let us now assume that the number fraction $x_\alpha$ of each conformation
$\alpha$ is fixed. Then, the fluid can be seen as a multicomponent
mixture of simple molecules. In this case the 
virial expansion can be written as \cite{HMD-06}
\be
Z = 1 - {c\over2} \sum_{\alpha,\beta} x_\alpha x_\beta I_{2,\alpha\beta} -
    {c^2\over3} \sum_{\alpha,\beta,\gamma} 
     x_\alpha x_\beta x_\gamma I_{3,\alpha\beta\gamma} + 
    O(c^3),
\label{expan-Z}
\end{equation}
where 
\begin{eqnarray}
I_{2,\alpha\beta} &\equiv & \int d^3{\bf r} f_{12} (r,\alpha\beta), \\
I_{3,\alpha\beta\gamma} &\equiv & \int d^3{\bf r}_{12} d^3{\bf r}_{13} \, 
       f_{12} (r_{12},\alpha\beta) f_{13} (r_{13},\alpha\gamma) 
       f_{23} (|{\bf r}_{12} - {\bf r}_{13}|,\beta\gamma),
\end{eqnarray}
and $c$ is the number density. 
If the molecules are rigid the number fraction $x_\alpha$ is 
density independent and equal to the zero-density probability $p_\alpha$.
Thus, if we define
\be
I_2 = \sum_{\alpha\beta} p_\alpha p_\beta I_{2,\alpha\beta} \qquad\qquad
I_3 = \sum_{\alpha\beta\gamma} 
      p_\alpha p_\beta p_\gamma I_{3,\alpha\beta\gamma},
\end{equation}
we obtain 
\be
Z = 1 - {c\over2} I_2 - {c^2\over3} I_3 + O(c^3),
\end{equation}
which is the usual virial expansion, with $T_1 = 0$. 
On the other hand, for flexible
molecules $x_\alpha$ is density dependent (if $x_\alpha$ were 
density-independent, single-molecule properties, for instance the 
radius of gyration, would not depend on density, which is clearly unphysical).
To derive the $c$ dependence of $x_\alpha$
we proceed as in ref.~\CITE{CMP-raggi}, Section II.
We consider a quantity $R^{(\alpha)}$ that assumes the value 1 if 
the conformation one is considering is the $\alpha$ one, and zero otherwise.
Explicitly, given a configuration $\beta$, the value $R^{(\alpha)}_\beta$ of
$R^{(\alpha)}$ on this configuration is
\be
R^{(\alpha)}_\beta = \delta_{\alpha\beta} = 
                     \cases{1 & if $\beta = \alpha$ \cr
                            0 & if $\beta \not= \alpha$ }.
\ee
By definition 
\be
  x_\alpha = \< R^{(\alpha)} \>.
\ee
In order to compute the virial expansion of the right-hand side, we consider
$L$ molecules in a volume $V$ and write
\be
  x_\alpha = {\sum_{\beta_1,\ldots \beta_L} 
 \int d{\mathbf r}_1\ldots d{\mathbf r}_L\, R^{(\alpha)}_{\beta_1}
  p_{\beta_1}\ldots p_{\beta_L} 
  \prod_{i<j} [1 + f_{ij}(|{\mathbf r}_i - {\mathbf r}_j|;\beta_i,\beta_j)] 
   \over 
 \sum_{\beta_1,\ldots \beta_L}
 \int d{\mathbf r}_1\ldots d{\mathbf r}_L \,
  p_{\beta_1}\ldots p_{\beta_L}
  \prod_{i<j} [1 + f_{ij}(|{\mathbf r}_i - {\mathbf r}_j|;\beta_i,\beta_j)] }.
\ee
Expanding the numerator we obtain 
\begin{eqnarray}
&& V^L \sum_{\beta_1} R^{(\alpha)}_{\beta_1} p_{\beta_1} + 
(L-1) V^{L-1} \sum_{\beta_1,\beta_2} R^{(\alpha)}_{\beta_1} p_{\beta_1}
p_{\beta_2} I_{2,\beta_1\beta_2} + \nonumber \\
&& \qquad 
{L-1\choose 2} V^{L-1} \sum_{\beta_1,\beta_2,\beta_3}
      R^{(\alpha)}_{\beta_1} p_{\beta_1} p_{\beta_2} p_{\beta_3}
      I_{2,\beta_2\beta_3} + \ldots \nonumber \\
&& = V^L p_\alpha + (L-1) V^{L-1} p_\alpha 
         \sum_{\beta} p_\beta I_{2,\alpha\beta} + 
      {L-1\choose 2} V^{L-1} p_\alpha I_2 + \ldots
\end{eqnarray}
Analogously, for the denominator we obtain
\be
V^L + {L\choose 2} V^{L-1} I_2 + \ldots
\ee
Then, in the thermodynamic limit, $L,V\to\infty$ at fixed $c \equiv L/V$, 
we obtain
\be
x_\alpha = p_\alpha + c p_\alpha \sum_\beta p_\beta I_{2,\alpha\beta} -
         c p_\alpha I_2 + O(c^2).
\label{xalpha}
\ee
In order to check the correctness of this result we have verified that
the condition $\sum x_\alpha = 1$ is satisfied by
our expression. If we now substitute this result for $x_\alpha$ into
Equation~(\ref{expan-Z}), we obtain for $B_3$
\be
B_3 = - {1\over 3} I_3^2 - \sum_{\alpha\beta\gamma} p_\alpha p_\beta p_\gamma
    I_{2,\alpha\beta} I_{2,\alpha\gamma} + I_2^2.
\ee
It is easy to recognize that 
\begin{eqnarray}
T_1 &=& 
 \sum_{\alpha\beta\gamma} 
   \int d^3{\mathbf r}_{12} d^3{\mathbf r}_{13}\,
   p_\alpha p_\beta p_\gamma f_{12}(r_{12},\alpha\beta)
       f_{13}(r_{13},\alpha\gamma) - I_2^2, \nonumber \\
   &=&
 \sum_{\alpha\beta\gamma} p_\alpha p_\beta p_\gamma
    I_{2,\alpha\beta} I_{2,\alpha\gamma} - I_2^2,
\end{eqnarray}
so that we reobtain (\ref{B3def}). Note that $T_1$ can also be written
as 
\be
T_1 = {1\over 2} \sum_{\alpha\beta} p_\alpha p_\beta 
   \left( \sum_\gamma p_\gamma I_{2,\alpha\gamma} - 
          \sum_\gamma p_\gamma I_{2,\beta\gamma} \right)^2.
\ee
This relation shows that $T_1\ge 0$ and that $T_1$ vanishes only if 
\be
\sum_\gamma p_\gamma I_{2,\alpha\gamma} = 
 \sum_\gamma p_\gamma I_{2,\beta\gamma}
\ee
for any pair $\alpha$ and $\beta$. Therefore, $T_1 = 0$ only if 
$\sum_\gamma p_\gamma I_{2,\alpha\gamma}$ is independent of $\alpha$. 
If this condition is satisfied, we have
$\sum_\gamma p_\gamma I_{2,\alpha\gamma} = I_2$, so that the density correction
that appears in Equation (\ref{xalpha}) vanishes. 
Therefore, $T_1$ vanishes only if 
$x_\alpha$ does not depend on $c$, that is if the molecules are rigid.

\section{Model and Simulation Details} \label{sec3}

We are
interested in determining universal properties in the good-solvent regime, 
in the 
limit in which the degree of polymerization $N$ goes to infinity.
We can thus use any model that captures the
basic polymer properties. For computational convenience we 
consider the well-known self-avoiding walk (SAW) model on a cubic lattice.
A star polymer with $f$ branches is represented by $f$
SAWs starting at a common point. A regular star 
is given by the center ${\bf r}_0$ and by  $f$ branches
${\bf r}_{i,j}$, $i=1,\ldots,f$, $j=1,\ldots,N_f$, such that
$|{\bf r}_0 - {\bf r}_{i,1}| = 1$, $|{\bf r}_{i,j} - {\bf r}_{i,j+1}| = 1$,
and all lattice sites are visited at most once. The total number of monomers
is $N = f N_f + 1$. This model is well defined up to $f = 6$. For larger
values of $f$, one should use a larger core or a model with 
soft interactions as in ref.~\CITE{HNG-04}.

\begin{figure}
\centerline{\epsfig{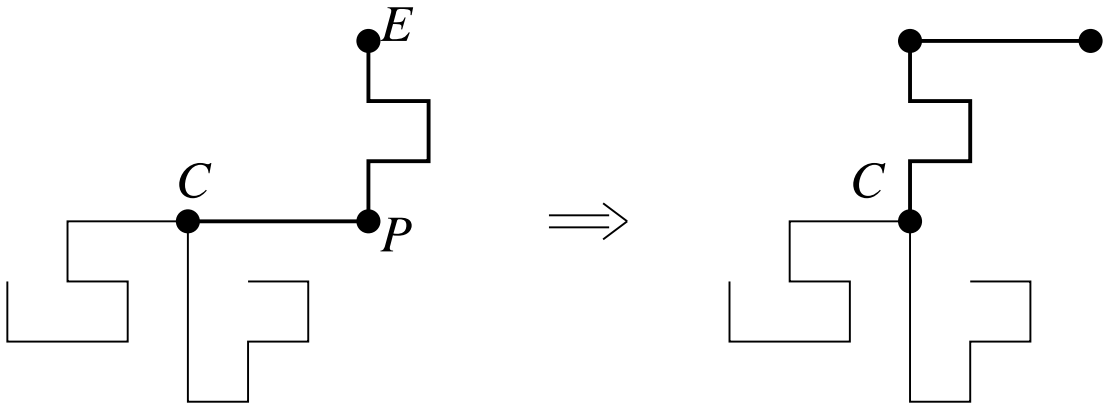}}
\vspace{1cm}
\caption{The cut-and-permute move applied to a star polymer with
$N_f = 8$ and $f = 3$. One first chooses an arm and a pivot point $P$.
The new arm is obtained by connecting the subwalk $PE$ to the center $C$
and then the subwalk $CP$ to the translated point $E$.
}
\label{fig:cp}
\end{figure}

We simulate the model by using different types of moves:
\begin{itemize} 
\item[(i)] We consider pivot moves\cite{Lal-69,MJHS-85,MS-88} applied to a 
single arm (analogous moves were used on the tetrahedral lattice 
in ref.~\CITE{Zifferer-99}). These moves have been shown to be
very efficient in simulations of linear polymers.\cite{MS-88}
In star-polymer simulations they are not equally efficient since
they are rarely accepted when the pivot is close to the
center of the star. 
\item[(ii)] We consider cut-and-permute moves\cite{Causo-02}
applied to a single arm (see Figure~\ref{fig:cp}).
They have been shown to be quite effective in simulations of polymers
grafted to impenetrable surfaces and speed up the conformational
changes close to the center of the star. 
\item[(iii)] We use local moves that involve moving one or two monomers 
of the walk.
\end{itemize}
For $f\le 5$ one can generalize the
arguments given in ref.~\CITE{MS-88} to show that this algorithm
is ergodic. No such proof is available for $f=6$, though we expect that
the combination of local and non-local moves makes the algorithm
ergodic in this case, too.

The virial coefficients are determined by using the hit-or-miss algorithm
discussed in refs.~\CITE{LMS-95,CMP-vir}.

\section{Results and Discussion} \label{sec4}

We perform simulations of star polymers for $f=4$ and $f = 6$ with 
$50 \le N_f \le 2000$ and $50 \le N_f \le 1000$ in the two cases,
respectively. Since $N = f N_f + 1$, the total
number of monomers is quite large; this should allow us to probe the universal
large-$N$ regime. Results for the constants $A_2$, $A_3$, and for 
$\widehat{A}_3 \equiv - {1\over3} I_3 \hat{R}_g^{-6}$ are reported in Table
\ref{tabledata}. In all cases the additional term $T_1$ gives a 
small negative contribution. Quantitatively we find $T_1/B_3\approx
0.065$, 0.05 for $f = 4,6$. This is consistent with the idea that 
star polymers are less and less flexible as $f$ increases, 
so that we expect $T_1/B_3\to 0$ as $f\to \infty$. 

\begin{table}
\caption{Estimates of $A_2$, $A_3$, and of
$\widehat{A}_3 \equiv  - {1\over3} I_3 \hat{R}_g^{-6}$.
}
\label{tabledata}
\begin{center}
\begin{tabular}{rrrrrrr}
\hline\hline
& \multicolumn{3}{c}{$f = 4$} & \multicolumn{3}{c}{$f = 6$} \\
$N_f$ & $A_2$ & $\widehat{A}_3$ & ${A}_3$ &
        $A_2$ & $\widehat{A}_3$ & ${A}_3$ \\ \hline
 50   & 10.899(2) & 52.48(4) & 49.75(3) & 15.280(2) & 113.43(6) & 108.79(6) \\
100   & 10.620(2) & 49.16(4) & 46.40(3) & 14.953(2) & 107.55(6) & 102.95(7) \\
150   & 10.497(2) & 47.64(4) & 44.98(5) & 14.809(2) & 105.10(6) & 100.33(6) \\
250   & 10.379(2) & 46.33(4) & 43.57(3) & 14.666(3) & 102.60(6) &  97.78(6) \\
500   & 10.257(2) & 44.98(4) & 42.24(4) & 14.515(3) &  99.99(6) &  95.45(6) \\
1000  & 10.173(2) & 44.13(4) & 41.48(5) & 14.416(3) &  98.38(8) &  93.72(8) \\
2000  & 10.115(2) & 43.47(4) & 40.82(4) & & \\
\hline\hline
\end{tabular}
\end{center}
\end{table}

\begin{figure}
\centerline{\epsfig{file=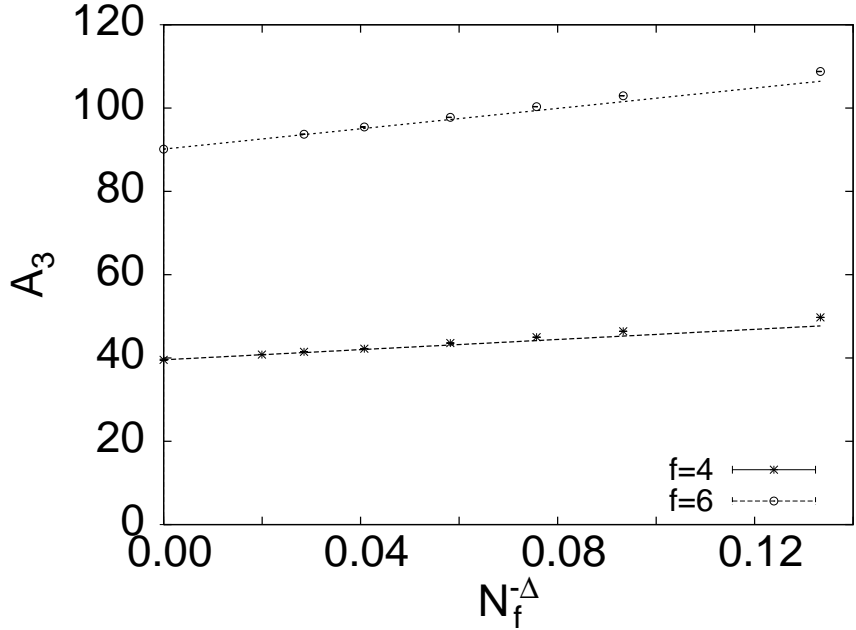,angle=-90,width=12truecm}}
\vspace{1cm}
\caption{Third virial invariant ratio $A_3$ for $f=4$ and $f=6$
versus $N^{-\Delta}$, $\Delta = 0.515$. We also report the extrapolation
function $A_3^* + a N_f^{-\Delta}$, $\Delta = 0.515$, determined in the
fit. }
\label{figA3}
\end{figure}

The data reported in Table \ref{tabledata} show a systematic dependence on 
$N_f$, see Figure~\ref{figA3}, 
and thus a proper extrapolation is needed in order to 
obtain the large-$N$ constants $A_2^*$ and $A_3^*$. We use the 
same procedure illustrated in ref.~\CITE{PH-05}, fitting the data to 
\be
A(N_f) = A^* + a N_f^{-\Delta} + b N_f^{-\Delta_2},
\ee
where $\Delta$ is a universal exponent whose best estimate is\cite{BN-97}
$\Delta = 0.515\pm 0.017$ (other results are reported in ref.~\CITE{PV-02}).
The exponent $\Delta_2$ 
is an effective one that takes into account several 
correction-to-scaling terms: as in ref.~\CITE{PH-05}, we take 
$\Delta_2 = 1.0\pm 0.1$. Of course, the previous expression is only
the leading part of an expansion in inverse (non-integer) powers of $N_f$. 
To monitor the role of the neglected terms, we have repeated the fit several
times, each time including only the data satisfying $N_f \ge N_{f,\rm min}$. 
Stable results for $A_2^*$ are obtained by using all data,
while good fits of $A_3$ require $N_{f,\rm min} = 100$. The corresponding 
results are:
\begin{eqnarray}
A_2^* &=& \cases{ 9.979 \pm 0.009 & $\qquad f = 4$, \cr 
                 14.174 \pm 0.016 & $\qquad f = 6$; }
\\[2mm]
A_3^* &=& \cases{ 39.56 \pm 0.16 & $\qquad f = 4$, \cr 
                  90.1  \pm 0.4  & $\qquad f = 6$. }
\end{eqnarray}
The errors we quote include the statistical uncertainty and the systematic
error due to the uncertainty on $\Delta$ and $\Delta_2$.
To compare with the literature it is useful to define the interpenetration
ratio $\Psi\equiv 2 (4\pi)^{-3/2} A_2$ and $g \equiv B_3/B_2^2 = A_3/A_2^2$.
We obtain
\begin{eqnarray}
\Psi^* &=& \cases{0.4480 \pm 0.0004 & $\qquad f = 4$, \cr 
                 0.6364 \pm 0.0007 & $\qquad f = 6$; }
\\[2mm]
g^* &=& \cases{ 0.397 \pm 0.002 & $\qquad f = 4$, \cr 
                0.449 \pm 0.002  & $\qquad f = 6$. }
\end{eqnarray}
For comparison we quote $\Psi^*$ and $g^*$ for $f=1$:\cite{CMP-vir}
\begin{eqnarray}
\Psi^* &=& 0.24693 \pm 0.00013, \\
g^* &=& 0.3240 \pm 0.0007. 
\end{eqnarray}
Other estimates of $\Psi^*$ and $g^*$ 
for $f = 1$ are quoted in refs.~\CITE{CMP-vir,PV-02,PV-07}.

There are no numerical results for $g^*$. Our estimates increase with $f$ 
as expected, but, for $f = 6$, $g^*$ is still far from the hard-sphere value
$5/8 = 0.625$, which should be valid for $f\to\infty$ 
(a discussion of the behavior of $g^*$ for large values of $f$ is reported in 
ref.~\CITE{RTD-95}). Our results for $\Psi^*$ are in reasonable agreement with 
the numerical
ones reported in the literature. For $f= 4$, refs.~\CITE{OSKK-96,RF-00,LK-02}
quote $\Psi^* = 0.46$, 0.467, $0.453\pm0.007$, while for $f = 6$ they
quote $\Psi^* = 0.64$, 0.665, $0.63\pm0.01$. The results of ref.~\CITE{RF-00}
are those that differ more significantly. Note, however, that in this work 
much smaller values of $N_f$ are used; moreover, no proper extrapolation is 
performed. Field theory results differ instead quite significantly,
predicting $\Psi^* = 0.517$, 0.798 for $f=4,6$, respectively.\cite{DF-84-2}
Recent experimental results for star polystyrene in benzene are 
reported in refs.~\CITE{ONNT-98,OINN-00}. They quote 
$0.43\lesssim \Psi \lesssim 0.46$ for $f = 4$ and 
$\Psi\approx 0.60$ for $f = 6$, in reasonable agreement with our
results. They also estimate the factor $g$. The 
results show a strong dependence on the molecular weight: the two samples with
highest molecular weight $M_w$ give $g \approx 0.40$, $g \approx 0.43$ for 
$f = 4$ and $g \approx 0.39$, $g \approx 0.50$ for $f = 6$, with 
$g$ increasing with $M_w$. 
These results are close to our estimates, even though the experimental
results apparently prefer somewhat larger values. Note that similar 
discrepancies are observed for linear polymers, see the experimental
results cited in refs.~\CITE{ONNT-98,OINN-00,OYY-04} and references therein.
Older results for $\Psi^*$ are cited in ref.~\CITE{DRF-90}. The experimental 
estimates of the 
interpenetration ratio for star polystyrene in toluene show a significant
dependence on the molecular weight $M_w$:
for the largest values of $M_w$ experiments give 
$\Psi\approx 0.46,0.55$ for $f = 4$ and $\Psi\approx 0.65$ for $f = 6$.
These results are reasonably close to our estimates. 
The experimental values of $\Psi$ for polybutadiene in cyclohexane 
quoted in ref.~\CITE{DRF-90}, $0.42\lesssim \Psi \lesssim 0.47$, 
are also consistent.

In conclusion, we have shown that $T_1$ is small but not negligible: if 
$T_1$ is neglected the error is of 6.5\% and 5\% for $f = 4,6$, respectively.
Moreover, our results allow us to compute the osmotic pressure in the
dilute regime with good precision (we expect the error to be of order 
of a few percent below the overlap concentration, see ref.~\CITE{CMP-vir}). 
We find:
\be
Z = \cases{ 1 + X + 0.397 X^2 + \ldots & $\qquad f = 4$, \cr
            1 + X + 0.449 X^2 + \ldots & $\qquad f = 6$,
          }
\ee
where $X \equiv B_2 c$.

\end{document}